\documentclass[aps,prl,reprint,superscriptaddress]{revtex4-2}
\usepackage{orcidlink}
\usepackage{graphicx}
\usepackage{dcolumn}
\usepackage{bm}
\usepackage{hyperref}
\usepackage[mathlines]{lineno}
\usepackage{amsmath}
\usepackage{subfigure}

\begin{document}



\title{First Observation of $\Lambda\pi^+$ and $\Lambda\pi^-$ Signals near the $\bar{K}N (I=1)$ Mass Threshold in $\Lambda_c^+\rightarrow\Lambda\pi^+\pi^+\pi^-$ Decay}

\noaffiliation
  \author{Y.~Ma\,\orcidlink{0000-0001-8412-8308}} 
  \author{J.~Yelton\,\orcidlink{0000-0001-8840-3346}} 
  \author{K.~Tanida\,\orcidlink{0000-0002-8255-3746}} 
  \author{I.~Adachi\,\orcidlink{0000-0003-2287-0173}} 
  \author{J.~K.~Ahn\,\orcidlink{0000-0002-5795-2243}} 
  \author{H.~Aihara\,\orcidlink{0000-0002-1907-5964}} 
  \author{S.~Al~Said\,\orcidlink{0000-0002-4895-3869}} 
  \author{D.~M.~Asner\,\orcidlink{0000-0002-1586-5790}} 
  \author{H.~Atmacan\,\orcidlink{0000-0003-2435-501X}} 
  \author{T.~Aushev\,\orcidlink{0000-0002-6347-7055}} 
% ...
\collaboration{The Belle Collaboration}



\begin{abstract}
  Using the data sample of 980 fb$^{-1}$ collected with the Belle detector operating at the KEKB asymmetric-energy $e^+e^-$ collider, we present the results of an investigation of the $\Lambda\pi^+$ and $\Lambda\pi^-$ invariant mass distributions looking for substructure in the decay $\Lambda_c^+\rightarrow\Lambda\pi^+\pi^+\pi^-$. We find a significant signal in each mass distribution. When interpreted as resonances, we find for the $\Lambda\pi^+$ ($\Lambda\pi^-$) combination a mass of
  $1434.3 \pm 0.6 (\mathrm{stat}) \pm 0.9(\mathrm{syst})$ MeV/$c^2$
  [$1438.5 \pm 0.9 (\mathrm{stat}) \pm 2.5(\mathrm{syst})$ MeV/$c^2$],
  an intrinsic width of
  $11.5 \pm 2.8 (\mathrm{stat}) \pm 5.3(\mathrm{syst})$ MeV/$c^2$
  [$33.0 \pm 7.5 (\mathrm{stat}) \pm 23.6(\mathrm{syst})$ MeV/$c^2$]  
  with a significance of 7.5$\sigma$ (6.2$\sigma$). As these two signals are very close to the $\bar{K}N$ threshold, we also investigate the possibility of a $\bar{K}N$ cusp, and find that we cannot discriminate between these two interpretations due to the limited size of the data sample.
\end{abstract}


\maketitle

Hyperon spectroscopy near the $\bar{K}N$ threshold has been a source of excitement for more than half a century.
There are many different scenarios that can generate states in this mass region.
The interplay of these scenarios makes the hyperon spectroscopy in this mass region particularly interesting.
A typical example is the $\Lambda(1405)(I=0)$ state,
which has been interpreted as an orbitally excited quark-diquark \cite{quark},
or as a $\bar{K}N$ bound state \cite{kn-molecue}.
On the other hand, the only known $I = 1$ state in this mass region is the $\Sigma(1385)$.
The standard quark model does not predict any more states near the $\Lambda(1405)$ mass, so if a $\Sigma^*$ resonance is observed, it may be exotic.
The $\bar{K}N (I=1)$ interaction is, most likely, not strong enough to produce a bound state, but a virtual state could exist \cite{oller} and could be observed as a threshold cusp.
The shape of such a cusp reflects the scattering length of the $\bar{K}N (I=1)$ interaction,
which is particularly interesting in relation with kaon condensation in neutron stars, where $K^-n$ interaction is most important.

\begin{figure}[h]
  \begin{center}
  \subfigure[]
  {\includegraphics[width=4.2cm]{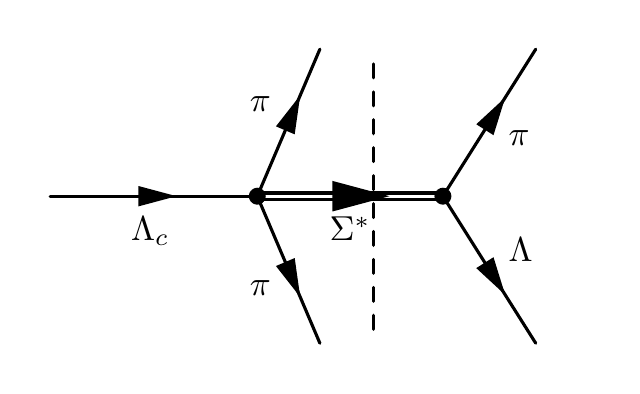}}
  \subfigure[]
  {\includegraphics[width=4.2cm]{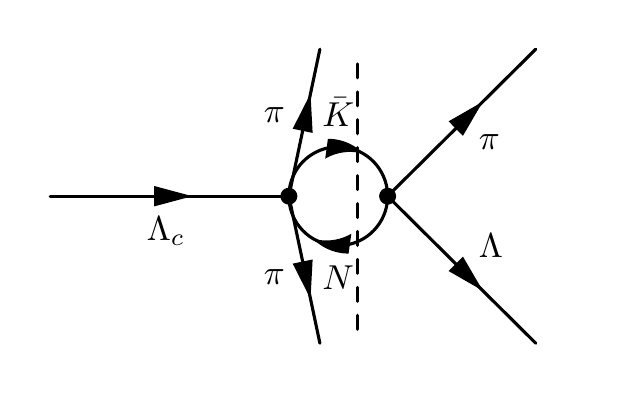}}
  \caption{
    $\Lambda_c^+ \to \Lambda \pi^+ \pi^- \pi^+$ decay as a virtual $\Lambda-\pi$ collider.
    The left side of the dashed line for both (a) and (b) contains an unknown $\Lambda_c$ decay form factor;
    the right side of the dashed line can be used to search for $\Sigma^*$ resonances (a) and
    study $\bar{K}N$ (re-)scattering with a cusp (b), respectively.
  }
  \label{diagram}
  \end{center}
\end{figure}

In this Letter we report a study of $\Lambda\pi^+$ and $\Lambda\pi^-$ invariant mass distributions
in the region above the $\Sigma(1385)$, in the decay $\Lambda_c^+\rightarrow\Lambda\pi^+\pi^+\pi^-$.   
The charge conjugate mode is implied throughout the current Letter.
Heavy baryon ($\Lambda_c^+$) decay provides a unique chance to investigate hyperon spectroscopy.
In particular, study of the $[\Lambda\pi^\pm]$ pair in the decay
$\Lambda_c^+ \to [\Lambda\pi^\pm] \pi^\mp \pi^+$ provides data comparable to a
$\Lambda-\pi$ collider in the range of 0 to 0.7 GeV/$c^2$ above the $\Lambda\pi^\pm$ mass threshold. This
allows analysis of the $I = 1$, $S = -1$ sector to be performed in this range. In the
present analysis, as shown in Fig. \ref{diagram}, we consider possible $\Sigma^*$ resonances,
and a $\bar{K}N$ threshold cusp.

Our data sample corresponds to the 980 fb$^{-1}$ integrated luminosity collected with the Belle detector at the KEKB
asymmetric-energy $e^+e^-$ collider \cite{kekb1,*kekb2}.
Most of the data are taken at the $\Upsilon(nS)$($n=1-5$) resonances
together with a small integrated luminosity collected off resonance.
Belle detector is a large-solid-angle magnetic spectrometer consisting of a silicon vertex detector (SVD),
a 50-layer central drift chamber (CDC), an array of aerogel threshold Cherenkov counters (ACC),
a barrel-like arrangement of time-of-flight scintillation counters (TOF),
and an electromagnetic calorimeter made of CsI(Tl) crystals (ECL).
These components are surrounded by a superconducting solenoid with a 1.5 T magnetic field.
The details of the Belle detector can be found in 
\cite{belle, *[{also see Section 2 in }]belle2}

We use a set of Monte Carlo simulation tools to optimize the event selections.
Event generation uses the EVTGEN \cite{evegen} package
and GEANT3 is used for Belle detector response simulation \cite{geant3}.
The event reconstruction starts with charged hadron ($h$) identification to select $p$ and $\pi^{\pm}$.
For each charged track, a likelihood [$\mathcal{L}$$(h)$] of particle identification (PID) is assigned
based on the measurement with CDC, TOF and ACC \cite{pid}.
The ratio of the PID likelihoods, $\mathcal{R}$$(h_S:h_B)$=$\mathcal{L}$$(h_S)$/[$\mathcal{L}$$(h_S)$+$\mathcal{L}$$(h_B)$],
is used for event selection.
Only $\pi^{\pm}$ with $\mathcal{R}$$(\pi:K)>0.2$ and $\mathcal{R}$$(\pi:p)>0.4$ and $p$ with $\mathcal{R}$$(p:\pi)>0.6$
are used in the following data analysis.
The PID efficiency is approximately $99\%$ for $\pi^\pm$ and $91\%$ for proton, respectively.
The kinetic information of the $\pi^{\pm}$ and $p$ obtained from the tracking device (SVD and CDC)
are used to reconstruct $\Lambda$ and $\Lambda_c^+$.
During the reconstruction,
daughter particles of the corresponding decay are fitted to the common vertex with the mass
of the mother particle as a mass-constraint vertex fit.
The so reconstructed $\Lambda$ trajectory is used for the $\Lambda_c^+$ reconstruction.
A confidence level $>0.001$ is required to select good $\Lambda$ and $\Lambda_c^+$ candidates.
To optimize the signal to noise ratio ($S/N$) of the final [$\Lambda\pi^{\pm}$] spectrum,
we require the scaled momentum of $\Lambda_c^+$ to satisfy $x_p = pc/\sqrt{s/4 - M_{\Lambda_c^+}^2c^4} > 0.43$,
where $p$ is the reconstructed $\Lambda_c^+$ momentum in the center-of-mass frame, $c$ is the speed of light,
$s$ is the square of the center-of-mass energy and $M_{\Lambda_c^+}$ is the mass of $\Lambda_c^+$.
A mass window of $\Delta M \le 8$ MeV/$c^2$ centered at the $\Lambda_c^+$ nominal mass 
corresponding to $\pm2.1\sigma$ of $\Lambda_c^+$ signal is also applied \cite{pdg}.
The $\Lambda_c^+$ decay vertex is required to satisfy $dr < 0.2$ cm and $dz < 2$ cm, where $dr$ and $dz$ are the distance from the
interaction point transverse to, and along, the $e^+$ beam direction.
In addition, to further improve the signal sensitivity in the $\Lambda\pi^+$ spectrum,
we veto $\Sigma(1385) \to \Lambda \pi^-$ contributions to the $\Lambda_c^+ \to \Lambda \pi^+ \pi^- \pi^+$ decay
with a mass window of 10 MeV centered at the nominal mass of $\Sigma(1385)^-$ \cite{pdg}.
This method is only effective for the $\Lambda\pi^+$ mode because of the different background level in the $\Lambda\pi^-$ mode.

\begin{figure}[h]
  \begin{center}
    \includegraphics[width=7cm]{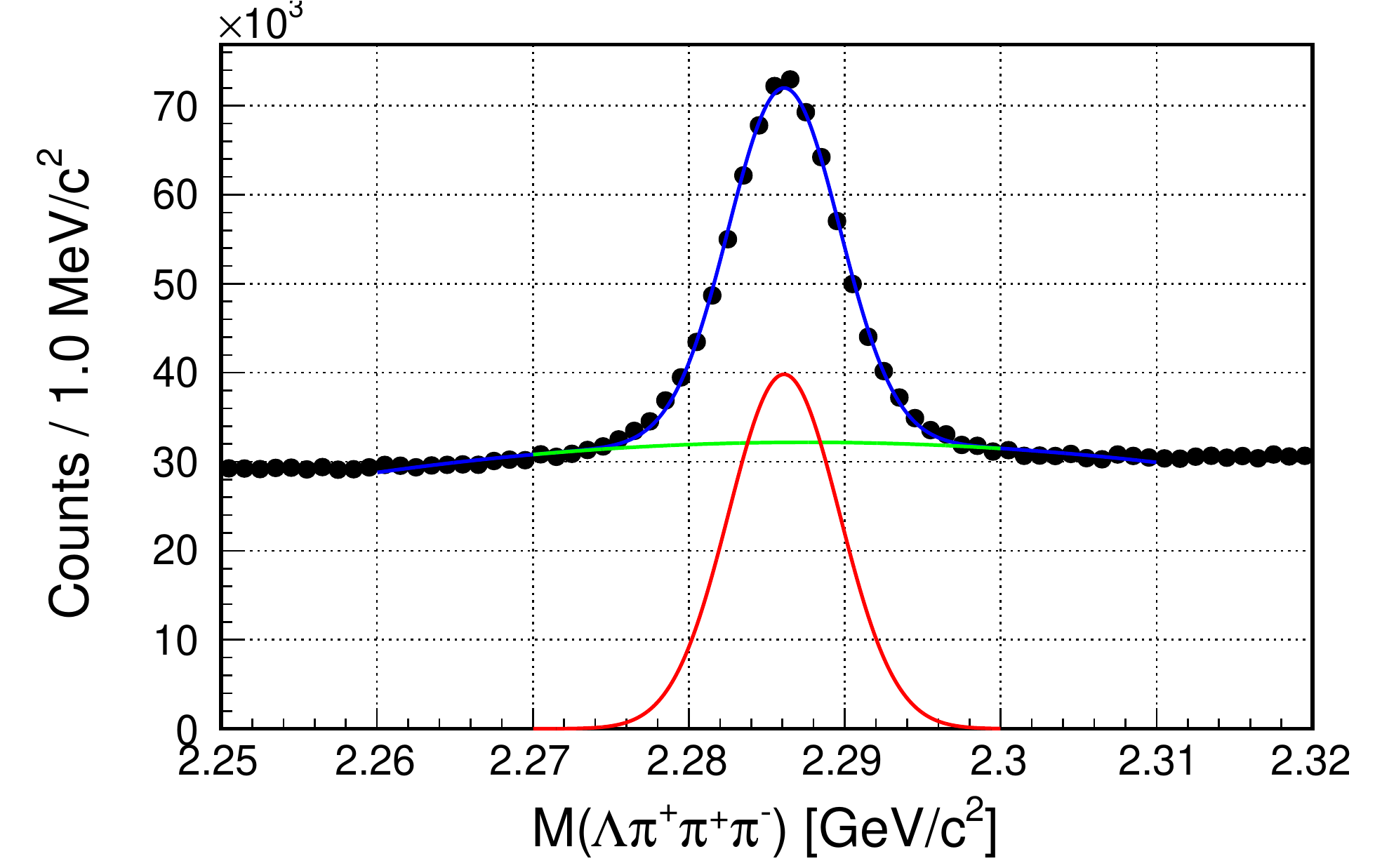}
    \caption{ The invariant mass of the $\Lambda \pi^+\pi^+\pi^-$ system after event selection.
      The green, red and blue curves represent the background, signal and total fit function, respectively.
      See text for details.}
    \label{fig:Lc_IM}
  \end{center}
\end{figure}

The reconstructed $\Lambda_c^+$ invariant mass from $\Lambda_c^+\rightarrow\Lambda\pi^+\pi^+\pi^-$ decay
after event selection is shown in Fig. \ref{fig:Lc_IM}.
The $\Lambda_c^+$ invariant mass is fitted with a second order Chebyshev polynomial for background
and a Gaussian function for signal.
The fitted $\Lambda_c^+$ invariant mass in our data analysis is $2286.12\pm 0.01$
and $0.34\pm0.14$ MeV/$c^2$ lower than the world average value,
where the uncertainty is dominated by the reference value \cite{pdg}.
We consider this $-0.34$ MeV/$c^2$ shift as the absolute Belle energy scale uncertainty
and will use it to estimate the systematic uncertainty for the $\Lambda\pi$ invariant mass.

The $[\Lambda\pi^{\pm}]$ invariant mass resolution is obtained by simulation and parametrized as a sum of two Gaussian functions
with $\sigma_1=1.15$ MeV/$c^2$, $\sigma_2=2.52$ MeV/$c^2$, an area ratio of 3.5, and a common mean.
The systematic uncertainty for the resolution function is estimated by comparing the reconstructed
$\Lambda_c^+$ invariant mass from simulation and Belle data.
The simulated resolution is $\sim4\%$ wider than the data,
which will be used to evaluate the systematic uncertainty in our data analysis.
In the following data analysis, the resolution function is convoluted
with the signal functions Eq. (\ref{eq:bw}) and Eq. (\ref{eq:cusp}) to derive the fit parameters of interest.

\begin{figure}[h]
  \vfill
  \subfigure[]
            {\includegraphics[width=7cm]{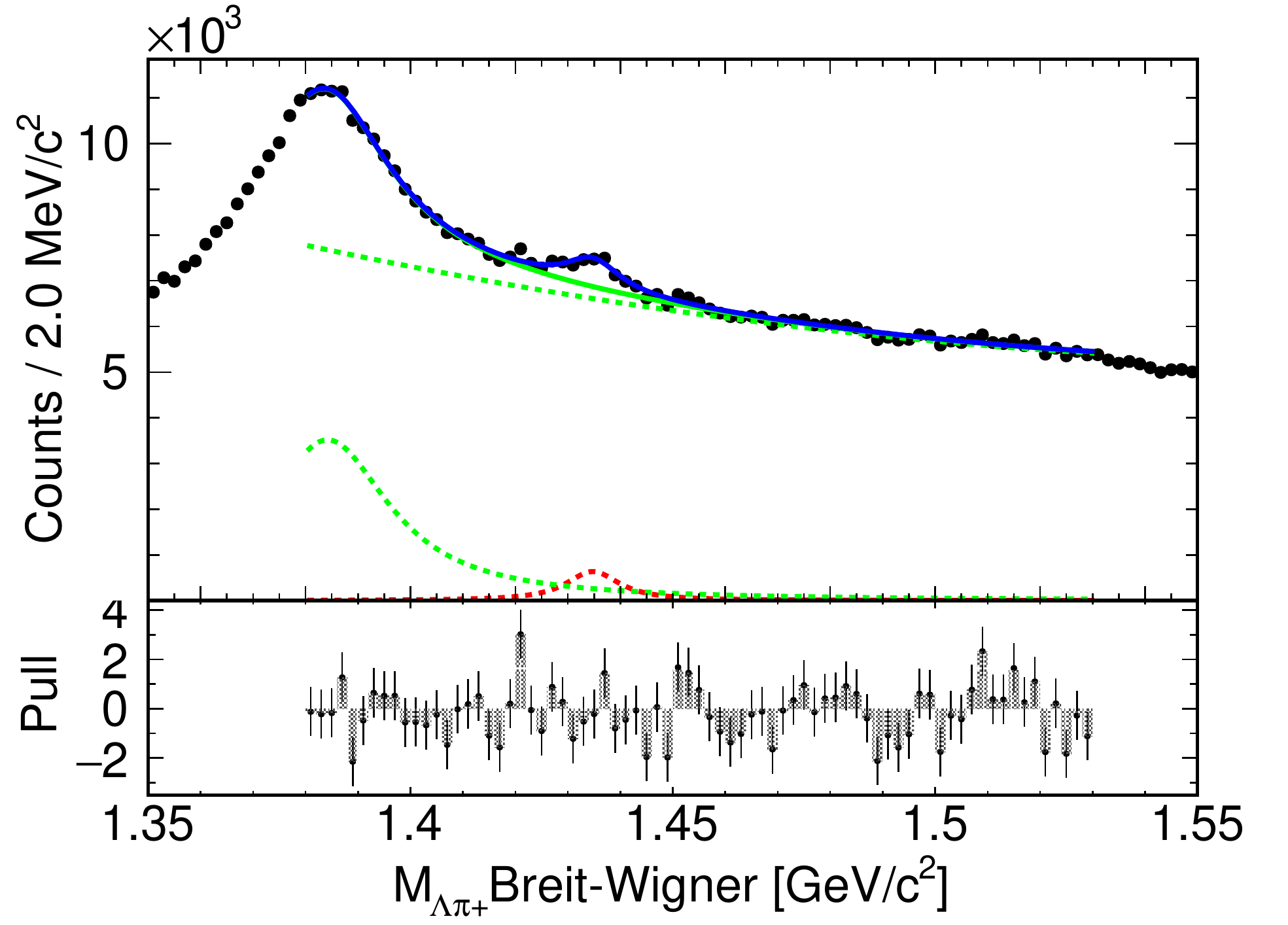}}
  \vfill
  \subfigure[]
            {\includegraphics[width=7cm]{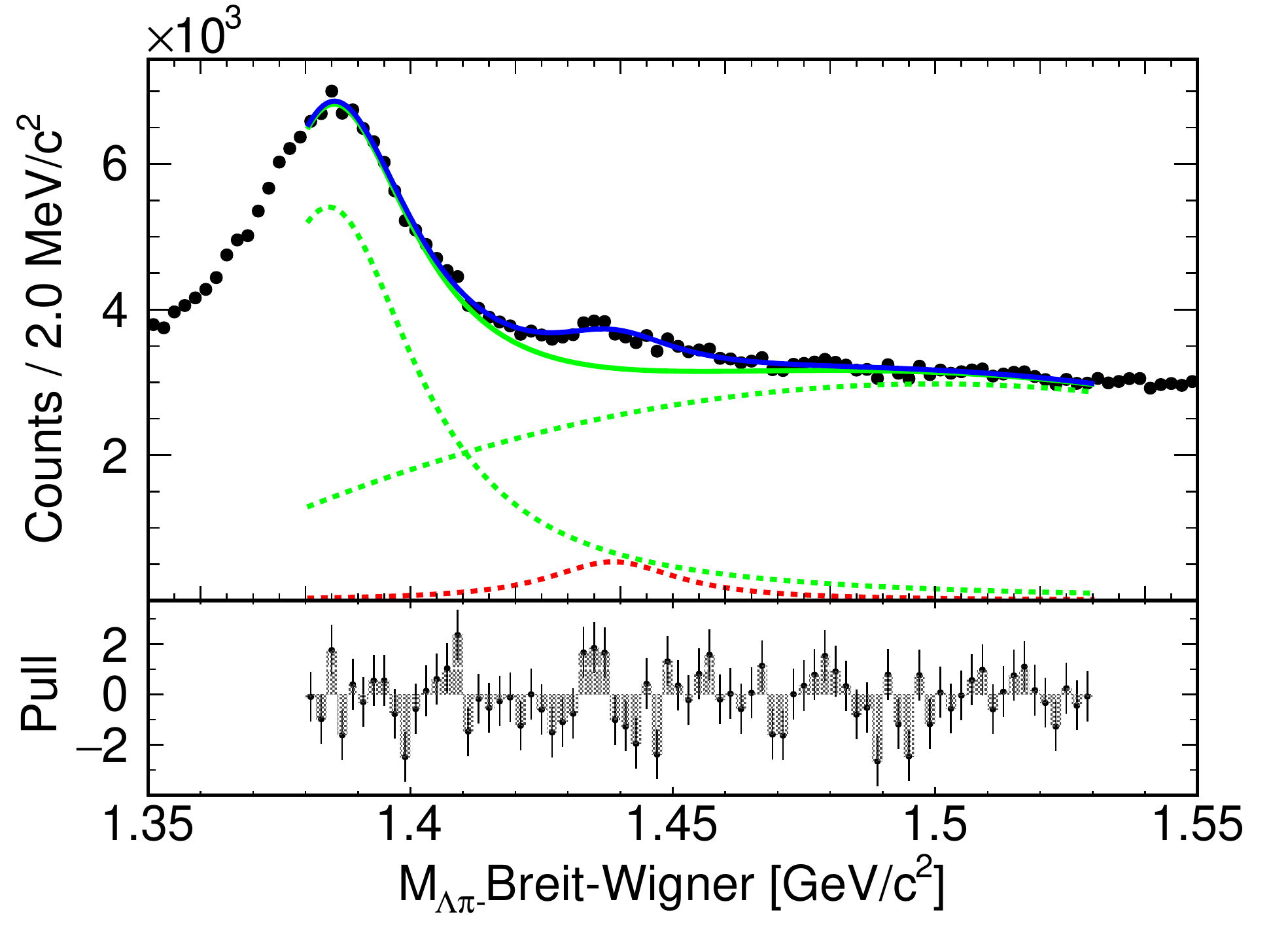}}
  \vfill
  \caption{$\Lambda\pi^+$ (a) and $\Lambda\pi^-$ (b) invariant mass distribution fitted with
    a Breit-Wigner signal and a background function.
    The solid (dotted) green, red and blue curves represent the (breakdown) background, signal and total fit function, respectively.
    Fitting range: 1.380$\sim$1.530 GeV/$c^2$. See text for details.}
  \label{bw_fit}
\end{figure}

\begin{figure}[h]
  \vfill
  \subfigure[c][]
            {\includegraphics[width=7cm]{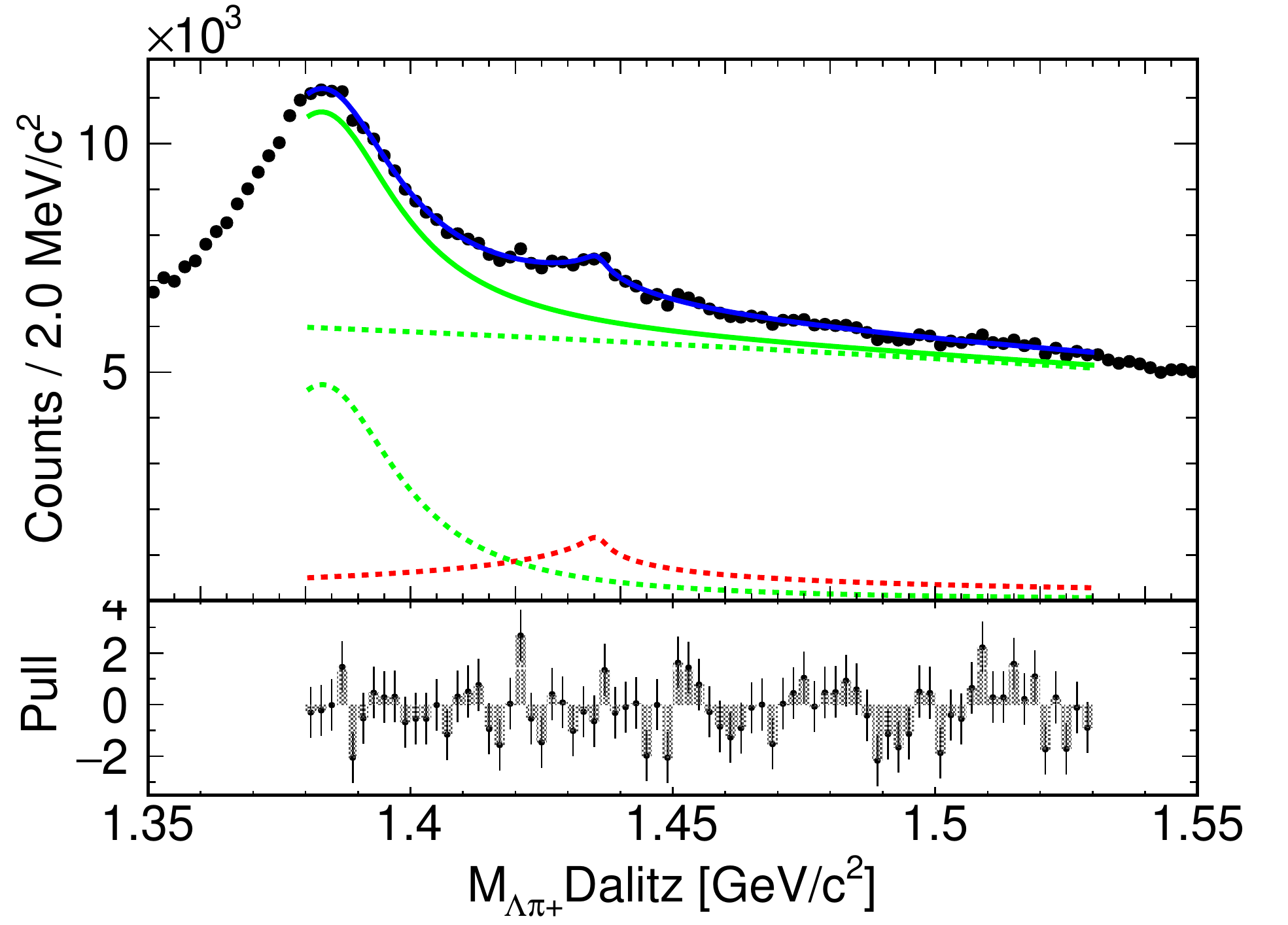}}
  \vfill
  \subfigure[]
            {\includegraphics[width=7cm]{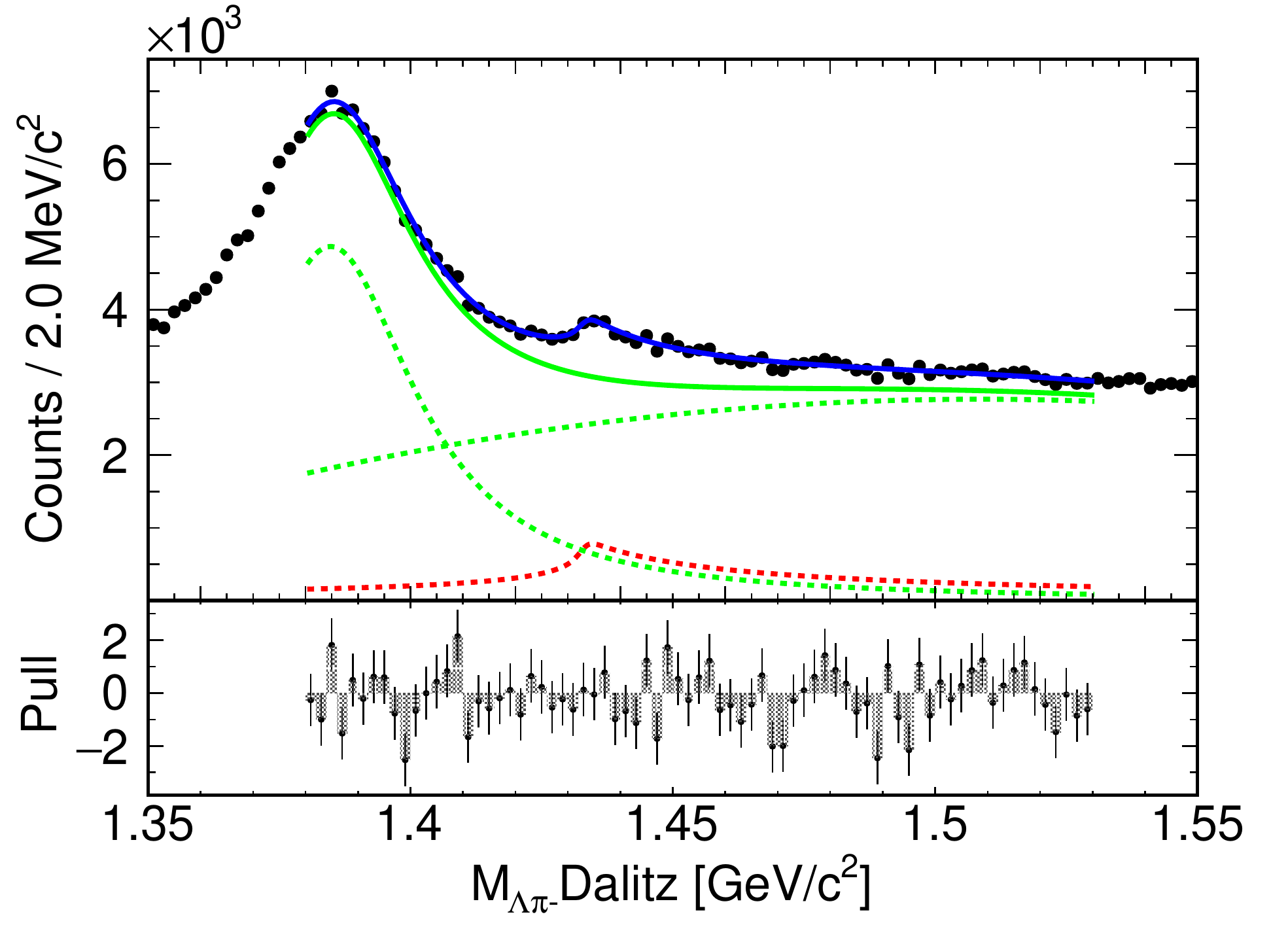}}
  \vfill
  \caption
      {$\Lambda\pi^+$ (a) and $\Lambda\pi^-$ (b) invariant mass distribution fitted with
        the Dalitz model for the signal and a background function.
        The solid (dotted) green, red and blue curves represent the (breakdown) background, signal and total fit function, respectively.
        Fitting range: 1.380$\sim$1.530 GeV/$c^2$. See text for details.}
\label{cusp_fit}
\end{figure}

The reconstructed [$\Lambda\pi^{\pm}$] invariant mass spectrum
from $\Lambda_c^+\rightarrow[\Lambda\pi^{\pm}]\pi^{\mp}\pi^+$ decay
after all event selections is shown in Figs. \ref{bw_fit} and \ref{cusp_fit}
for two different fitting models as will be explained in the text.
The reconstruction efficiency is evaluated to be $\sim 8\%$ based on simulation.
Clear enhancements near the $\bar{K}N$ mass thresholds are observed in the
$\Lambda\pi^+$ and $\Lambda\pi^-$ invariant mass spectrum, respectively.
We have confirmed that these enhancements originate from $\Lambda_c^+$ decays
by comparing the $\Lambda\pi^\pm$ invariant mass spectrum
from the $\Lambda_c^+$ mass window and the side band regions
centered at $\Lambda_c^+$ nominal mass $\pm 5 \sigma$ with $\pm 1.1 \sigma$ width.

We investigate the signals using two different
parametrizations of the signal shape: a Breit-Wigner
function which describes a $\Sigma^*$ resonance, and the Dalitz model \cite{dalitz} which describes a $\bar{K}N$ cusp.
A common background function for both $\Lambda\pi^\pm$ charge modes is used.
It consists of two components: a Breit-Wigner function for the $\Sigma(1385)^\pm$ contribution and
a second-order Chebyshev polynomial function for the high-mass background events.
Five free parameters, i.e., relative yield between these two components, peak and width of the Breit-Wigner function
and two coefficients of the Chebyshev polynomial function are used for the fit.

To interpret the signals as $\Sigma^*$ resonances, we use a nonrelativistic Breit-Wigner function defined as
\begin{flalign}
  \label{eq:bw}
  f_\text{BW} = \frac{\Gamma/2}{(E-E_\text{BW})^2+\Gamma^2/4},
\end{flalign}
where $E$ is the $\Lambda\pi^{\pm}$ invariant mass, $E_\text{BW}$ is the $\Sigma^*$ mass, and $\Gamma$ is the resonance width.
A binned least-$\chi^2$ fit to the spectrum is shown in Fig. \ref{bw_fit} and the fit results are summarized in Table \ref{tab:bw_fit}.

\begin{table}[h]
  \begin{center}
    \caption{Breit-Wigner fitting results. The quoted errors are statistical only.}
    \begin{tabular}{ >{\centering\arraybackslash} m{0.12\linewidth}  >{\centering\arraybackslash} m{0.3\linewidth} 
        >{\centering\arraybackslash} m{0.2\linewidth} 
        >{\centering\arraybackslash} m{0.2\linewidth} }
      \hline
      \hline
          {\tt Mode} & $E_\text{BW}$ [MeV/$c^2$] & $\Gamma$ [MeV/$c^2$] & $\chi^2$ / NDF \\
          \hline
          $\Lambda\pi^+$ & $1434.3 \pm 0.6$ & $11.5 \pm 2.8$ & $74.4 / 68$ \\ 
          \hline
          $\Lambda\pi^-$ & $1438.5 \pm 0.9$ & $33.0 \pm 7.5$ & $92.3 / 68$ \\           
          \hline
    \end{tabular}
    \label{tab:bw_fit}
  \end{center}
\end{table}

Given the overlap of the observed signals and the $\bar{K}N$ mass threshold,
it is natural to expect a strong $\bar{K}N$ contribution via $\bar{K}$-$N$ rescattering
as illustrated in Fig. \ref{diagram} (b).
In particular, by neglecting the $\Lambda_c^+$ decay form factor, the $\bar{K}N$ cusp can be related to
the $\bar{K}$-$N$ complex scattering length ($A=a+ib$) with the Dalitz model \cite{dalitz}:
\begin{flalign}
  \label{eq:cusp}
  f_{D} &= \frac{4\pi b}{(1+kb)^2 + (ka)^2}, \textrm{$E > m_{\bar{K}N}$} \nonumber\\
  &= \frac{4\pi b}{(1+\kappa a)^2 + (\kappa b)^2}, \textrm{$E < m_{\bar{K}N}$},
\end{flalign}
where 
$k$ and $\kappa$ are the magnitude of the $\bar{K}N$ relative momentum above and below threshold, respectively.
Specifically,
$k=\sqrt{2\mu(E - m_{\bar{K}N})}$ for $E>m_{\bar{K}N}$ and
$\kappa=\sqrt{2\mu(m_{\bar{K}N} - E)}$ for $E<m_{\bar{K}N}$, where $\mu = 1/(1/m_{N} + 1/m_{\bar{K}})$ is the reduced mass of the $\bar{K}N$ system.

Figure \ref{cusp_fit} shows the fit result with the Dalitz model by using binned least-$\chi^2$ method.
The obtained scattering length is given in Table \ref{tab:cusp_fit}.
The quoted uncertainties are statistical only.

\begin{table}[h]
  \begin{center}
    \caption{Dalitz model fitting results.
    }
    \begin{tabular}{ >{\centering\arraybackslash} m{0.2\linewidth}  >{\centering\arraybackslash} m{0.2\linewidth} 
        >{\centering\arraybackslash} m{0.2\linewidth} 
        >{\centering\arraybackslash} m{0.2\linewidth} }
      \hline
      \hline
          {\tt Mode} & $a [\mathrm{fm}]$ & $b [\mathrm{fm}]$ & $\chi^2$ / NDF \\
          \hline
          $\Lambda\pi^+$ & $0.48 \pm 0.32$ & $1.22 \pm 0.83$ & $68.9 / 68$ \\
          \hline
          $\Lambda\pi^-$ & $1.24 \pm 0.57$ & $0.18 \pm 0.13$ & $78.1 / 68$ \\           
          \hline
    \end{tabular}
    \label{tab:cusp_fit}
  \end{center}
\end{table}

To test the interpretation of the observed signal as $\bar{K}N$ cusps,
we adopt the Flatt\'{e} parametrization given in \cite{baru}:
\begin{flalign}
  \label{eq:flatte}
  &f_{Fl}&& \nonumber\\
  &= \frac{\Gamma}{(E - m_{\bar{K}N} - E_\text{BW})^2+(\Gamma+gk)^2/4},\textrm{$E > m_{\bar{K}N}$} &&\nonumber\\ 
  &= \frac{\Gamma}{(E - m_{\bar{K}N} - E_\text{BW} - g\kappa/2)^2+\Gamma^2/4}, \textrm{$E < m_{\bar{K}N}$},&& 
\end{flalign}
where $E$, $E_\text{BW}$, and $\Gamma$ have the same definition as in Eq. (\ref{eq:bw}),
$g$ is the coupling constant to $\bar{K}N$,
and $k$ ($\kappa$) is the magnitude of the $\bar{K}N$ relative momentum above (below) threshold as defined in Eq. (\ref{eq:cusp}).
However, it is known that the three parameters ($E_\text{BW}, \Gamma$, and $g$) in Flatt\'{e} parametrization are not independent,
but are correlated by the so-called scaling behavior \cite{baru}.

The scaling behavior suggests that
by fixing $E_\text{BW}$ far away from the ${\bar{K}N}$ mass threshold ($\sim$ 1.435 GeV/$c^2$),
the Flatt\'{e} parametrization effectively describes a cusp.
This can be made explicit with the help of the effective range expansion,
which relates the Flatt\'{e} parametrization to the scattering lengths as \cite{baru}
\begin{equation}
  A = a +ib = \frac{g}{2(E_\text{BW} - i\Gamma/2)}.
  \label{eq11}
\end{equation}
This relation is based on the process of $\bar{K}$-$N$ rescattering into the $\Lambda\pi$ final state as illustrated in Fig. \ref{diagram} (b).
In particular, we can customize the Flatt\'{e} parametrization by
replacing $\Gamma = 2E_\text{BW}b/a$,
$g = 2E_\text{BW}(a + b^2/a) / \hbar c$ and set an arbitrarily large $E_\text{BW}$ ($10$ GeV was used in this analysis).
The scattering length derived this way is largely consistent with the Dalitz model.
The difference is summarized in Table \ref{tab:sys_err_cusp} as a systematic uncertainty.


In the Breit-Wigner fit, three types of systematic uncertainties are considered:
uncertainty due to the absolute Belle energy scale,
uncertainty induced by the resolution function
and uncertainty due to the fitting procedure.
The absolute Belle energy scale is estimated to be $0.34$ MeV/$c^2$ lower than the reference value as previously described.
We therefore shift the $\Lambda\pi^{\pm}$ invariant mass spectrum by $+0.34$ MeV/$c^2$
to derive the systematic uncertainty.
The uncertainty induced by the resolution function is checked by shrinking the resolution by $4\%$,
which is obtained by comparing $\Lambda_c^+$ mass distribution between simulation and data as mentioned before.
To evaluate the uncertainty due to the fitting procedure,
we define a new background function as a third-order Chebyshev polynomial and change the fitting range to
1.420 $\sim$ 1.520 GeV/$c^2$ for both $\Lambda\pi$ charge modes.
By excluding $\Sigma(1385)^\pm$ from the fitting range,
we estimate the uncertainty related to the $\Sigma(1385)^\pm$ background component.
The resulting systematic uncertainties for the Breit-Wigner fitting are summarized in Table \ref{tab:sys_err_bw},
where the independent contributions are added in quadrature for the total.
To be conservative, we take the larger value of any uncertainty
that is asymmetric and use it as a symmetric uncertainty for the final result.

For the Dalitz model fit, we also consider
the absolute Belle energy scale uncertainty,
resolution uncertainty and the fitting procedure uncertainty,
where the same treatment as the Breit-Wigner case is used.
In addition, we also include the uncertainty induced by the data fitting model, 
which is obtained by comparing the difference between the Dalitz model and Flatt\'{e} parametrizations.
The systematic uncertainties are summarized in Table \ref{tab:sys_err_cusp},
where the independent contributions are added in quadrature for the total.
Similar to the Breit-Wigner case, we take the larger value of any uncertainty
that is asymmetric and use it as a symmetric uncertainty for the final result.

\begin{table}[h]
  \begin{center}
    \caption{Systematic uncertainties for the Breit-Wigner fitting parameters in [MeV].}
    \begin{tabular}{ >{\centering\arraybackslash} m{0.2\linewidth}  >{\centering\arraybackslash} m{0.12\linewidth}  >{\centering\arraybackslash} m{0.3\linewidth} 
        >{\centering\arraybackslash} m{0.2\linewidth} }
      \hline
      \hline
          Source & {\tt Mode} & $E_\text{BW}$ [MeV/$c^2$] & $\Gamma$ [MeV/$c^2$]\\
          Energy scale & $\Lambda\pi^+$ & $+0.5$ & $-0.1$ \\
          Resolution & $\Lambda\pi^+$ & $0.0$ & $+0.1$ \\
          Fitting procedure & $\Lambda\pi^+$ & $+0.8$ & $-5.3$ \\
          \hline
          \noalign{\vskip 3pt}
          Total & $\Lambda\pi^+$ & $^{+0.9}_{-0.0}$ & $^{+0.1}_{-5.3}$ \\
          Final value & & $\pm{0.9}$ & $\pm{5.3}$ \\
          \noalign{\vskip 3pt}
          \hline
          Energy scale & $\Lambda\pi^-$ & $+0.2$ & $-2.9$ \\
          Resolution & $\Lambda\pi^-$ & $0.0$ & $+0.1$ \\
          Fitting procedure & $\Lambda\pi^-$ & $-2.5$ & $-23.4$ \\
          \hline
          \noalign{\vskip 3pt}
          Total & $\Lambda\pi^-$ & $^{+0.2}_{-2.5}$ & $^{+0.1}_{-23.6}$ \\
          Final value & & $\pm{2.5}$ & $\pm{23.6}$ \\
          \noalign{\vskip 3pt}
          \hline
    \end{tabular}
    \label{tab:sys_err_bw}
  \end{center}
\end{table}

\begin{table}[h]
  \begin{center}
    \caption{Systematic uncertainties for the Dalitz fitting parameters in [fm].}
    \begin{tabular}{ >{\centering\arraybackslash} m{0.2\linewidth}  >{\centering\arraybackslash} m{0.12\linewidth}  >{\centering\arraybackslash} m{0.2\linewidth} 
        >{\centering\arraybackslash} m{0.2\linewidth} }
      \hline
      \hline
          Source & {\tt Mode} & $a$ [fm] & $b$ [fm] \\
          Energy scale & $\Lambda\pi^+$ & $+0.03$ & $-0.17$ \\
          Resolution & $\Lambda\pi^+$ & $-0.01$ & $-0.03$ \\
          Fitting procedure & $\Lambda\pi^+$ & $+0.37$ & $+2.54$ \\
          Model & $\Lambda\pi^+$ & $+0.05$ & $-0.05$ \\
          \hline
          \noalign{\vskip 3pt}
          Total & $\Lambda\pi^+$ & $^{+0.38}_{-0.01}$ & $^{+2.54}_{-0.18}$ \\
          Final value & & $\pm{0.38}$ & $\pm{2.54}$ \\
          \noalign{\vskip 3pt}
          \hline
          Energy scale & $\Lambda\pi^-$ & $+0.19$ & $-0.10$ \\
          Resolution & $\Lambda\pi^-$ & $-0.01$ & $0.00$ \\
          Fitting procedure & $\Lambda\pi^-$ & $+1.55$ & $-0.17$ \\
          Model & $\Lambda\pi^-$ & $-0.16$ & $-0.05$ \\          
          \hline
          \noalign{\vskip 3pt}
          Total & $\Lambda\pi^-$ & $^{+1.56}_{-0.16}$ & $^{+0.00}_{-0.20}$ \\
          Final value & & $\pm{1.56}$ & $\pm{0.20}$ \\
          \noalign{\vskip 3pt}
          \hline
    \end{tabular}
    \label{tab:sys_err_cusp}
  \end{center}
\end{table}

The statistical significance of the signals is derived by excluding the peaks from the fit,
finding the change in the log-likelihood ($\Delta[\ln(L)]$) and converting this to a $p$ value taking into account the
change in the number of degrees in freedom.
This is then converted to an effective number of standard deviations.
The same treatment is applied for the two fitting procedures and for each $\Lambda\pi$ charge mode.
The lowest significance out of four combinations is reported in this Letter as
7.5$\sigma$ for the $\Lambda\pi^+$ mode and 6.2$\sigma$ for the $\Lambda\pi^-$ mode, respectively.


To understand the $\Lambda\pi^{\pm}$ invariant mass enhancement,
two interpretations as $\Sigma^*$ resonances and $\bar{K}N$ threshold cusps are tested.
For the $\Sigma^*$ interpretation, we have the final results for
$\Lambda\pi^+$ ($\Lambda\pi^-$) peak of $M$ =
$1434.3 \pm 0.6 (\mathrm{stat}) \pm 0.9(\mathrm{syst})$ MeV/$c^2$
[$1438.5 \pm 0.9 (\mathrm{stat}) \pm 2.5(\mathrm{syst})$ MeV/$c^2$],
width =
$11.5 \pm 2.8 (\mathrm{stat}) \pm 5.3(\mathrm{syst})$ MeV/$c^2$
[$33.0 \pm 7.5 (\mathrm{stat}) \pm 23.6(\mathrm{syst})$ MeV/$c^2$].
It is noticeable that the $\Lambda\pi^+$ peak $E_\text{BW} = 1433.9 \pm 0.6$ MeV/$c^2$ is lower than the
$\bar{K}^0p$ mass threshold $m_{\bar{K}^0p} = 1435.9$ MeV/$c^2$
and the $\Lambda\pi^-$ peak $E_\text{BW} = 1437.7 \pm 0.9$ MeV/$c^2$ is higher than the
$K^-n$ mass threshold $m_{K^-n} = 1433.2$ MeV/$c^2$,
where the reference values for the kaon and nucleon mass are taken from world average \cite{pdg}.

If we interpret the signals as threshold cusps, we can derive the following scattering lengths:
$\bar{K}^0$-$p$ ($K^-$-$n$) with $\Lambda\pi^+$ ($\Lambda\pi^-$) mode as
$a = 0.48 \pm 0.32 (\mathrm{stat}) \pm 0.38(\mathrm{syst})$ fm
[$ 1.24 \pm 0.57 (\mathrm{stat}) \pm 1.56(\mathrm{syst})$ fm]
and
$b = 1.22 \pm 0.83 (\mathrm{stat}) \pm 2.54(\mathrm{syst})$ fm
[$ 0.18 \pm 0.13 (\mathrm{stat}) \pm 0.20(\mathrm{syst})$ fm].
The scattering length derived in our data analysis is larger than the previous results \cite{kp, ikeda, yamagata}.
This difference may be due to the neglected decay form factor of an order of 0.5 fm.

On the theoretical side,
both interpretations are discussed, but the cusp interpretation may be more favored.
Oller and Mei{\ss}ner discussed the possibility of a resonance in the 
$I=1$ channel and reported a pole at $(1444.0-i69.4)$ MeV in the second
Riemann sheet \cite{oller}. However, the imaginary part is too large to explain 
the present structure. Many theories \cite{ikeda, kamiya, yamagata} predicted
a threshold cusp in the $I=1$ channel. In addition, Ref. \cite{oller} also
reported another (virtual) pole in the third Riemann sheet below the threshold,
which could produce a cusp at the threshold.



We report the first observations of $\Lambda\pi^{\pm}$ invariant mass enhancements near the $\bar{K}N$ mass thresholds
in the substructure of the $\Lambda_c^+\rightarrow\Lambda\pi^+\pi^+\pi^-$ decay. 
The significance for the observed signals are 7.5$\sigma$ for $\Lambda\pi^+$ combination,
and 6.2$\sigma$ for $\Lambda\pi^-$ combination, respectively.
Limited by the statistics and the shape of the background,
we cannot distinguish between $\Sigma^*$ resonances and $\bar{K}N$ threshold cusps,
since both fits give similar $\chi^2$s.

\begin{acknowledgments}

This work, based on data collected using the Belle detector, which was
operated until June 2010, was supported by 
the Ministry of Education, Culture, Sports, Science, and
Technology (MEXT) of Japan, the Japan Society for the 
Promotion of Science (JSPS), and the Tau-Lepton Physics 
Research Center of Nagoya University; 
the Australian Research Council including Grants
No.~DP210101900, 
No.~DP210102831, 
No.~DE220100462, 
No.~LE210100098, 
No.~LE230100085; 
Austrian Federal Ministry of Education, Science and Research (FWF) and
FWF Austrian Science Fund No.~P~31361-N36;
the National Natural Science Foundation of China under Contracts
No.~11675166,  
No.~11705209;  
No.~11975076;  
No.~12135005;  
No.~12175041;  
No.~12161141008; 
Key Research Program of Frontier Sciences, Chinese Academy of Sciences (CAS), Grant No.~QYZDJ-SSW-SLH011; 
Project ZR2022JQ02 supported by Shandong Provincial Natural Science Foundation;
the Ministry of Education, Youth and Sports of the Czech
Republic under Contract No.~LTT17020;
the Czech Science Foundation Grant No. 22-18469S;
Horizon 2020 ERC Advanced Grant No.~884719 and ERC Starting Grant No.~947006 "InterLeptons" (European Union);
the Carl Zeiss Foundation, the Deutsche Forschungsgemeinschaft, the
Excellence Cluster Universe, and the VolkswagenStiftung;
the Department of Atomic Energy (Project Identification No. RTI 4002) and the Department of Science and Technology of India; 
the Istituto Nazionale di Fisica Nucleare of Italy; 
National Research Foundation (NRF) of Korea Grants
No.~2016R1\-D1A1B\-02012900, No.~2018R1\-A2B\-3003643,
No.~2018R1\-A6A1A\-06024970, No.~RS\-2022\-00197659,
No.~2019R1\-I1A3A\-01058933, No.~2021R1\-A6A1A\-03043957,
No.~2021R1\-F1A\-1060423, No.~2021R1\-F1A\-1064008, No.~2022R1\-A2C\-1003993;
Radiation Science Research Institute, Foreign Large-size Research Facility Application Supporting project, the Global Science Experimental Data Hub Center of the Korea Institute of Science and Technology Information and KREONET/GLORIAD;
the Polish Ministry of Science and Higher Education and 
the National Science Center;
the Ministry of Science and Higher Education of the Russian Federation, Agreement 14.W03.31.0026, 
and the HSE University Basic Research Program, Moscow; 
University of Tabuk research Grants
No.~S-1440-0321, No.~S-0256-1438, and No.~S-0280-1439 (Saudi Arabia);
the Slovenian Research Agency Grant No. J1-9124 and No.~P1-0135;
Ikerbasque, Basque Foundation for Science, Spain;
the Swiss National Science Foundation; 
the Ministry of Education and the Ministry of Science and Technology of Taiwan;
and the United States Department of Energy and the National Science Foundation.
These acknowledgements are not to be interpreted as an endorsement of any
statement made by any of our institutes, funding agencies, governments, or
their representatives.
We thank the KEKB group for the excellent operation of the
accelerator; the KEK cryogenics group for the efficient
operation of the solenoid; and the KEK computer group and the Pacific Northwest National
Laboratory (PNNL) Environmental Molecular Sciences Laboratory (EMSL)
computing group for strong computing support; and the National
Institute of Informatics, and Science Information NETwork 6 (SINET6) for
valuable network support.

\end{acknowledgments}

\bibliography{Y1435_PRL.bib}
\bibliographystyle{apsrev4-2}

\end{document}